\documentclass[12pt]{article}
\usepackage{amsmath}
\usepackage{amsthm}
\usepackage{amssymb}
\usepackage{amsfonts}
\usepackage{epic}
\usepackage{eepic}
 \usepackage{times}
\usepackage{mathrsfs}   
\usepackage[matrix,arrow]{xy}

\theoremstyle{plain}

\newtheorem{thm}{Theorem}
\newtheorem{propn}{Proposition}
\newtheorem{lem}{Lemma}

\theoremstyle{remark}
\newtheorem{rem}{Remark}

\newcommand\beq{\begin{equation}}
\newcommand\enq{\end{equation}}
\newcommand\bem{\begin{multline}}
\newcommand\enm{\end{multline}}

\def\beqa{\begin{eqnarray}}
\def\eeqa{\end{eqnarray}}
\def\ba{\begin{array}}
\def\ea{\end{array}}

\def\det{\operatorname{det}}

\newcommand{\f}[2]{{\ensuremath{%
    \mathchoice%
    {\dfrac{#1}{#2}}
    {\dfrac{#1}{#2}}
    {\frac{#1}{#2}}
    {\frac{#1}{#2}}
}}}
\newcommand{\tf}[2]{\ensuremath{#1/#2}}
\newcommand{\pa}[1]{\ensuremath{\left(#1\right)}}
\newcommand{\paa}[1]{\ensuremath{\left\{#1\right\}}}
\newcommand{\pac}[1]{\ensuremath{\left[#1\right]}}
\newcommand{\paf}[2]{\ensuremath{\left(\f{#1}{#2}\right)}}

\let\tend=\rightarrow

\def\ga{\gamma}
\def\Ga{\Gamma}
\def\de{\delta}

\def\la{\lambda}

\def\th{\theta}

\def\vth{\vartheta}

\def\ua{\uparrow}
\def\da{\downarrow}

\newcommand{\mc}[1]{\ensuremath{\mathcal{#1}}}

\newcommand{\msc}[1]{\ensuremath{\mathscr{#1}}}

\newcommand{\bs}[1]{\ensuremath{\boldsymbol{#1}}}

\newcommand{\ov}[1]{\ensuremath{\overline{#1}}}
\newcommand{\wt}[1]{\ensuremath{\widetilde{#1}}}

\newcommand{\Int}[2]{\ensuremath{\int\limits_{#1}^{#2}}}

\newcommand{\sul}[2]{\ensuremath{\sum\limits_{#1}^{#2}}}
\newcommand{\pl}[2]{\ensuremath{\prod\limits_{#1}^{#2}}}

\newcommand{\R}{\ensuremath{\mathbb{R}}}
\newcommand{\Cx}{\ensuremath{\mathbb{C}}}

\newcommand{\Dp}[1]{\ensuremath{\partial_{#1}}}

\newcommand{\limit}[2]{\ensuremath{\underset{#1 \tend #2}{\longrightarrow} }}

\newcommand{\ex}[1]{\ensuremath{\e{e}^{#1}}}

\newcommand{\ddet}[2]{\ensuremath{\det_{#1}\pac{#2}}}

\newcommand{\abs}[1]{\ensuremath{\left| #1 \right|}}

\newcommand{\norm}[1]{\ensuremath{\left\|#1\right\|}}

\newcommand{\dd}{\mathrm{d}}
\newcommand{\e}[1]{\ensuremath{\mathrm{#1}}}

\newcommand{\intn}[2]{\ensuremath{[\![ \, #1 \,;\, #2 \,]\!]}}

\newcommand{\vt}{\vartheta}

\newcommand{\eeq}{\end{equation}}

\newcommand{\CH}{{\mathcal H}}

\newcommand{\CW}{{\mathcal W}}

\newcommand{\BR}{{\mathbb R}}

\newcommand{\BS}{{\mathbb S}}

\newcommand{\rf}[1]{(\ref{#1})}

\newcommand{\aufz}
{\begin{list}{$\bullet$}{\topsep0cm \itemsep0cm \parsep0cm}}
\newcommand{\eaufz}{\end{list}}

\begin{document}


\title{TBA for the Toda chain}

\author{K. K. Kozlowski, J. Teschner}


\maketitle

\centerline{\bf Abstract}
\begin{quote}
{\small 
We give a direct derivation of a proposal of Nekrasov-Shatashvili 
concerning the quantization conditions of the Toda chain.
The quantization conditions are formulated in terms 
of solutions to a nonlinear integral equation similar
to the ones coming from the thermodynamic Bethe ansatz.
This is equivalent to extremizing a certain function
called Yang's potential. It is shown that the 
Nekrasov-Shatashvili formulation of  the quantization conditions
follows from the solution theory of the Baxter equation,
suggesting that this way of formulating the quantization 
conditions should indeed be applicable to large classes of
quantized algebraically integrable models.}
\end{quote}

\section{Introduction}
\setcounter{equation}{0}

The $N$-body quantum mechanical Hamiltonian
\beq
{\bf H}=  \sul{\ell=1}{N} \f{ \bs{p}^2_{\ell}}{2} + \pa{\kappa g^2}^{\hbar}\ex{\bs{x}_N-\bs{x}_1} +\sul{k=2}{N} g^{2\hbar} \ex{\bs{x}_{k-1}-\bs{x}_k}  \;.
\enq
is known as the quantum Toda chain. Above,  $\bs{p}_{\ell}$ and $\bs{x}_k$ are quantum observables satisfying the canonical commutation relations
$\pac{\bs{p}_{\ell},\bs{x}_k}=i\hbar \de_{\ell,k}$, $g$ and $\kappa$ are coupling constant. When $\kappa=1$
one deals with the so-called closed Toda chain and, 
when $\kappa=0$, with the open Toda chain.

This model appears to be a prototype for
an interesting class of integrable models called
algebraically integrable models.  It  
was introduced and solved, on the classical level, by Toda 
\cite{TodaClassicalSolutionTodaChain} in 1967. Then, in 1977, Olshanetsky and 
Perelomov
\cite{OlshanetskyPerelomovIntegralsOfMotionSemi-SimpleLieAlgebraSystems} constructed the set of $N$ commuting and independent integrals of motion for the 
closed chain, thus proving the so-called
quantum integrability of the model. In 1980-81, Gutzwiller \cite{GutzwillerTodaChainSmallNEigenfunctionsI} was 
able to build explicitly the eigenfunctions and write down the quantization conditions 
for small numbers of particles ($N=2,3,4$). In particular he expressed the
eigenfunctions of the closed chain with $N$-sites
as a linear combination of the eigenfunctions of the open chain with $N-1$ particles. 

A particularly succesful approach to the solution of the quantum Toda chain
was initiated by Sklyanin. In 1985, Sklyanin applied the 
quantum inverse scattering method 
(QISM) to the study of the Toda chain. This led to
the development of the so-called quantum separation of variables method. 
In this novel framework, 
he was able to obtain the Baxter equations for the model, 
from which Gutzwiller's quantization conditions can be obtained.
The Baxter equation was later re-derived
by Pasquier and Gaudin \cite{GaudinPasquierTQOperatorTodaChain} with the help
of an explicit construction of the so-called Q-operator, similar to 
the method developed by Baxter for the solution of the eight 
vertex model \cite{BaxterExactlySolvableLatticeModels}.

In '99 Kharchev and Lebedev 
\cite{KharchevLebedevTodaChainEigenfunctions}
constructed the multiple integral 
representations for the eigenfunctions of the closed $N$-particle
Toda chain. Their construction can be seen as a generalization of 
Gutzwiller's solution allowing one to express the 
eigenfunctions of the 
closed $N$-particle Toda chain in terms of those of the 
open chain with $N-1$ particles for all $N$. 
In '09, An \cite{AnCompletenessofEigenbasisToda} completed the picture by 
proving  rigorously that Gutzwiller's quantization conditions 
are necessary and sufficient for obtaining a state in the spectrum.

However, the form of the quantization conditions 
obtained in the above-mentioned works appears to be
rather involved.
Recently Nekrasov and Shatashvili proposed in  
\cite{NekrasovShatashviliConjectureTBADescriptionSpectrumIntModels} 
that the quantization conditions for the Toda chain can
be reformulated in terms of the solutions 
to a nonlinear integral equation (NLIE) similar to the equations
originating in the thermodynamic Bethe ansatz method.
With the help of the solutions to the relevant nonlinear integral
equation, Nekrasov and Shatashvili defined a function $\CW$ whose
critical points are in a one-to-one correspondence with the
simultaneous eigenstates of the conserved quantitites.
This formulation not only seems to be in some respects more  
efficient than the
previous one, it also indicates an amazing universality 
of the form the quantization conditions may take in 
integrable models.

The proposal of 
\cite{NekrasovShatashviliConjectureTBADescriptionSpectrumIntModels}  
was based on rather indirect arguments coming from the 
study of supersymmetric gauge theories. It seems desirable 
to derive the proposal more directly from the integrable structure
of the model. Our main aim in this note is to give such 
a derivation. It is obtained from the solution 
theory of the Baxter equation. In other integrable models there 
are known connections between the Baxter 
equation and nonlinear integral equations that look similar to the one
that appears here, see e.g. \cite{BLZ,Za,Te}.
However, the precise form of the NLIE depends heavily on the
analytic properties that the relevant solutions of the 
Baxter equation must have in the different models.
In the present case of a particle system we encounter an 
interesting new feature: the quantization conditions are 
not formulated as equations on the zeros of the
solutions of the Baxter equation, but instead, they 
are equations on the poles of the so-called quantum Wronskian
formed from two linearly independent solutions of the 
Baxter equation. The positions $\de=(\de_1,\dots,\de_N)$ 
of these poles are the
variables that the Yang's potential $\CW=\CW(\de)$ depends on.

It is worth stressing that
the Baxter equation or generalizations thereof have a good 
chance to figure as a universal tool for the study of the
spectrum of quantum integrable models. Our method of derivation
strongly indicates that similar formulations of the 
quantization conditions should 
exist for large classes of quantized
{\it algebraically} integrable models.

This article is organized as follows. We first recall 
how the Separation of Variables method reduces
the problem to find the eigenstates of the Toda chain
to the problem to find a certain set of solutions to
the Baxter functional equation specified by strong 
conditions on the analyticity and the asymptotics of 
its elements.
Then, in Section  \ref{Sec:q-cond} we explain how  
Gutzwiller's quantization conditions can be reformulated
in terms of the solutions to a certain NLIE and in terms
of the Yang's potential $\CW(\de)$. The derivation
of this reformulation is sketched. 
The proofs of our claims are presented in the Appendices.
In Appendix \ref{Section NLIE and spectrum}, we establish 
the relevant properties of Gutzwiller's basis of fundamental 
solutions to the T-Q equation. Then in Appendix 
\ref{existence/uniqueness} we prove the existence and 
uniqueness of solutions to the 
NLIE introduced in Section  \ref{Sec:q-cond}. 
In Appendix
\ref{TBA vs. BAX} we give rigorous proofs of 
the main results presented in 
Section  \ref{Sec:q-cond}.

{\small {\bf Acknowledgements.}
The authors gratefully acknowledge support
from the EC by the Marie Curie Excellence
Grant MEXT-CT-2006-042695.

We are happy to dedicate this paper to T. Miwa on the 
occasion of his 60th birthday.
}

\section{Separation of variables approach to the Toda chain}
\setcounter{equation}{0}

\subsection{Integrability of the Toda chain}

The integrability of the Toda chain follows from the existence of Lax matrices
\beq
L_n\pa{\la}= \bigg(\,\ba{cc}  \la - \bs{p}_n  &
        g^{\hbar} \,\ex{-\bs{x}_n} \\
        - g^{\hbar}\, \ex{\bs{x}_n}  & 0  \ea   \bigg)
\qquad [\,\bs{x}_n\,,\,
\bs{p}_n\,]\,=\,i \;,
\enq
satisfying a Yang-Baxter equation with a rational, six-vertex type, R-matrix. Thus, the set of $\kappa$-twisted monodromy matrices
\beq
M\pa{\la} = \bigg(
\ba{cc} 1 & 0 \\ 0 & \kappa^{\hbar} \ea\bigg) 
L_N\pa{\la} \dots L_1\pa{\la} \;.
\enq
allows one to build the transfer matrix ${\mathbf T}\pa{\la}=\e{tr}\pac{M\pa{\la}}$ which is the generating function of the set of $N$ commuting Hamiltonians
associated with the Toda chain
\beq
{\mathbf T}\pa{\la} =  \la^N + \sul{k=0}{N-1} \pa{-1}^k \la^{N-k} {\bf H}_k\; .
\enq
The first two Hamiltonians read
\beq
 {\bf H}_1 = \sul{k=1}{N} \bs{p}_k = {\bf P} \qquad
{\bf H}_2=  \f{ {\bf P}^2}{2} - \paa{ \sul{\ell=1}{N} \f{\bs{p}^2_{\ell}}{2} + g^{2\hbar}\kappa^{\hbar}\ex{\bs{x}_N-\bs{x}_1} +\sul{k=2}{N} g^{2\hbar} \ex{\bs{x}_{k-1}-\bs{x}_k} } \;.
\enq
Any eigenvector of the transfer matrix defines a polynomial 
$\bs{t}\pa{\la}=\prod_{k=1}^{N}\pa{\la-\tau_k}$.
The $N$ commuting Hamiltonians are self-adjoint, hence the set $\paa{\tau}$ is necessarily self conjugated: 
$\paa{\tau_k}=\paa{\ov{\tau}_k}$.

\subsection{Separation of variables}\label{SOV}

The Separation of Variables (SOV) method was developed for the Toda chain 
in \cite{SklyaninSoVFirstIntroTodaChain,GaudinPasquierTQOperatorTodaChain,KharchevLebedevTodaChainEigenfunctions,
AnCompletenessofEigenbasisToda}. The main results of these works
may be summarized as follows:

The wave-functions $\Psi_{\bs{t}}^{}(x)$,  $x=(x_1,\dots,x_N)$ of any eigenstate
to the transfer matrix ${\mathbf T}(\la)$ with eigenvalue $\bs{t}(\la)$ 
can be represented 
by means of an integral transformation of the form 
\begin{equation}\label{SOVtransf}
\Psi_{\bs{t}}^{}(x)\,=\,\int_{\BR^{N-1}} \dd\mu(\ga)\;\Phi_{\bs{t}}^{}(\ga)\,\Xi_P(\ga|x)\,,
\end{equation} 
where integration is over vectors $\ga=(\ga_1,\dots,\ga_{N-1})\in\BR^{N-1}$
with respect to a measure $\dd \mu(\ga)$ first found in
\cite{SklyaninSoVFirstIntroTodaChain},
$\Xi_P(\ga|x)$ is an integral kernel for which the explicit expression 
can be found in \cite{KharchevLebedevTodaChainEigenfunctions},
$P$ is the eigenvalue of the center of mass momentum ${\mathbf P}$ in the
state $\Psi_{\bs{t}}^{}$, and $\Phi_{\bs{t}}^{}(\ga)$ is the wave-function in the
so-called SOV-representation. The key feature of the SOV representation
is that $\Phi_{\bs{t}}^{}(\ga)$ takes a factorized form 
\begin{equation}\label{SOVfactor}
\Phi_{\bs{t}}^{}(\ga)\,=\,\prod_{k=1}^{N-1}q_{\bs{t}}^{}(\ga_k)\,.
\end{equation}
The integral transformation \rf{SOVtransf} is constructed in such a way
that the eigenvalue equation for the family of operators
${\mathbf T}(\la)$ is equivalent to the fact that the function 
$q_{\bs{t}}^{}(y)$ which represents the state $\Psi_{\bs{t}}$ via \rf{SOVtransf} and
\rf{SOVfactor} satisfies the so-called Baxter equation,
\begin{equation}\label{BAX}
\bs{t}(\la)q_{\bs{t}}^{}(\la)\,=\,i^Ng^{N\hbar}q_{\bs{t}}^{}(\la+i\hbar) +
\kappa^{\hbar}(-i)^Ng^{N\hbar}q_{\bs{t}}^{}(\la-i\hbar)\,.
\end{equation}
The integral transformation \rf{SOVtransf} can be inverted to express
$\Phi_{\bs{t}}^{}(\ga)$ in terms of $\Psi_{\bs{t}}^{}(x)$.
In this way it becomes possible to find the necessary and sufficient
conditions that $q_{\bs{t}}^{}(y)$ has to satisfy in 
order to represent an eigenstate of ${\mathbf T}(\la)$
via \rf{SOVtransf} and
\rf{SOVfactor}. The conditions are 
\cite{KharchevLebedevTodaChainEigenfunctions,
AnCompletenessofEigenbasisToda}
\begin{align*}
{\rm (i)} \quad & \text{$\bs{t}(\la)$ is a polynomial of the form $\textstyle \bs{t}(\la)=\prod_{k=1}^N(\la-\tau_k)$, with $\paa{\tau_k}=\paa{\ov{\tau}_k}$.}\\
{\rm (ii)} \quad & \text{$q(\la)$ is entire and has asymptotic behavior
$\abs{ q\pa{\la} }=\e{O}\big( \ex{-\f{N\pi}{2\hbar}\abs{\Re\pa{\la}}}  \abs{\la}^{\f{N}{2\hbar}\pa{ 2\abs{\Im\pa{\la}}-\hbar } } \big)$}.
%
\end{align*}
Above, the $\e{O}$ symbol is uniform in the strip $\paa{z\; : \; \abs{\Im\pa{z}}\leq \tf{\hbar}{2}}$. This reduces the problem to construct
all the eigenfunctions and finding the complete spectrum of the Toda chain
to finding the set $\BS$ of all 
solutions $\pa{\bs{t}\pa{\la}, q_{\bs{t}}^{}\pa{\la}}$ 
to the Baxter equation \rf{BAX}
that satisfy the conditions (i) and (ii) above.

\section{Quantization conditions}\label{Sec:q-cond}
\setcounter{equation}{0}

It turns out that the Baxter equation \rf{BAX} admits solutions within
the class $\BS$ described above only for a discrete set of choices for the 
polyonomial $\bs{t}(\la)$. This is what expresses the quantization of the
spectrum of ${\mathbf T}(\la)$ within the 
SOV-framework. Our first aim in this section will 
be to outline how to reformulate 
the resulting conditions on $\bs{t}(\la)$ more concretely, following the
approaches initiated by Gutzwiller and Pasquier-Gaudin.

It gives useful insight to divide the problem to construct and classify 
the solutions to the Baxter equation  \rf{BAX} which satisfy (i) and (ii) 
into in two steps. In the first step, one weakens the analytic requirements
(ii) on $q(\la)$ slightly by allowing $q(\la)$ to have a certain number
of poles. In this case, it will be possible to find 
two linearly independent solutions $q_{\bs{t}}^\pm(\la)$ 
to \rf{BAX} for arbitrary $\bs{t}(\la)$ satisfying (i). 
In the second step, one constructs the solution $q(\la)$ 
satisfying (i) and (ii) in the form
\begin{equation}\label{ansatz}
q(\la)\,=\,P_+q_{\bs{t}}^+(\la)+P_-q_{\bs{t}}^-(\la)  \,,
\end{equation}
where $P_\pm$ are constants.
The requirement that $q(\la)$ is entire means that the poles of 
$q_{\bs{t}}^\pm(\la)$ must cancel each other in \rf{ansatz} which is
only possible if $\bs{t}(\la)$ is fine-tuned in a suitable way.
This is the origin of the quantization of the spectrum
of ${\mathbf T}(\la)$.

\subsection{Gutzwiller's formulation of the quantization conditions}
 
It turns out that there is a canonical minmal choice for the set
of poles of $q_{\bs{t}}^\pm(\la)$ that one needs to allow. 
One needs to allow $N$ poles $\de_1,\dots,\de_N$ whose 
positions are determined by the choice of $\bs{t}(\la)$. 
More precisely, out of $\bs{t}(\la)$ one constructs the so-called Hill determinant
\beq
\mc{H}\pa{\la} = \det \pac{ \ba{cccccc}  \dots &  \ddots & \ddots& \ddots & \dots & \dots\\
                                        \dots & \f{\rho^{\hbar}}{\bs{t}\pa{\la-i\hbar}} & 1 & \f{1}{\bs{t}\pa{\la-i\hbar}}& 0 & \dots \\
        \dots  &0 & \f{\rho^{\hbar}}{\bs{t}\pa{\la}} & 1 & \f{1}{\bs{t}\pa{\la}}& 0 \dots \\
        \ddots & \ddots  & \ddots & \ddots & \ddots  &\ddots
  \ea} \; ,
\label{definition determinant infini Hill}
\enq
where $\rho:=\kappa g^{2N}$. 
It can be shown that $\CH(\la)$ admits the representation 
\beq
\mc{H}\pa{\la} = \pl{a=1}{N} \f{  \sinh\f{\pi}{\hbar}\pa{\la-\de_a}  }{  \sinh\f{\pi}{\hbar}\pa{\la-\tau_a} } \; , 
\label{equation factorisation H}
\enq
where one chooses $\abs{\Im\pa{\de_k}}<\tf{\hbar}{2}$. This defines $\de_1,\dots,\de_N$ in terms of $\bs{t}(\la)$.

Let us then, instead of $\BS$ consider the class $\BS'$ of
solutions to \rf{BAX} which satisfy the conditions (i) and
$({\rm ii})'$, 
\begin{align*}
 {\rm ({ii})'} \quad & \text{$q(\la)$ is meromorphic with set of poles
contained in $\{\de_1,\dots,\de_N\}$ 
and}\\
& \text{it has an asymptotic behavior
$\abs{ q\pa{\la} }=\e{O}\big( \ex{-\f{N\pi}{2\hbar}\abs{\Re\pa{\la}}}  \abs{\la}^{\f{N}{2\hbar}\pa{ 2\abs{\Im\pa{\la}}-\hbar } } \big)$}.
\end{align*}
The Baxter equation \rf{BAX} has two linearly independent 
solutions $q_{\bs{t}}^\pm(\la)$
within $\BS'$ for arbitrary $\bs{t}(\la)$. 
One possible construction of the
solutions $q_{\bs{t}}^\pm(\la)$ goes back to Gutzwiller's work on the Toda chain.
They may be defined as follows:
\begin{equation}\label{qfromQ}
q_{\bs{t}}^\pm(\la)\,=\,\frac{Q_{\bs{t}}^\pm(\la)}{\prod_{a=1}^N \big\{\ex{-\f{\pi \la}{\hbar}}\sinh\frac{\pi}{\hbar}(\la-\de_a) \big\}}\,,
\end{equation}
where 
\beq\label{Qpmdefinition}
 Q_{\bs{t}}^+(\la)\,=\, \f{ \pa{\kappa g^N}^{-i\la} K_+\pa{\la}  \ex{-N\f{\pi}{\hbar}\la} }{\pl{k=1}{N}\hbar^{-i\frac{\la}{\hbar}} \Ga(1-i(\la-\tau_k)/\hbar)} \,, 
\quad
 Q_{\bs{t}}^-(\la)\,=\,\f{  g^{iN\la} K_-\pa{\la}  \ex{-N\f{\pi}{\hbar}\la}  }{\pl{k=1}{N}\hbar^{i\frac{\la}{\hbar}} \Ga(1+i(\la-\tau_k)/\hbar)}   \,.
\enq
with $K_{\pm}\pa{\la}$ being half-infinite determinants:
\beq
K_+\pa{\la} = \det \pac{ \ba{cccccc}  1  & \bs{t}^{-1} \pa{\la+i\hbar} & 0 & \cdots    \\
            \f{\rho^{\hbar}}{\bs{t}\pa{\la+2i\hbar} } & 1 &    \bs{t}^{-1}\pa{\la+2i\hbar} & 0 & \cdots \\
            0 & \ddots & \ddots & \ddots  &   \ddots  & \cdots
                                    \ea }
\label{definition determinant K+}
\enq
and $K_-\pa{\la}=\ov{K_+\pa{\ov{\la}}}$ (recall that $\paa{\tau_k}=\paa{\ov{\tau}_k}$). 
%
%
%
%
%
%
%
%
For the reader's convenience we have included a self-contained proof
that $q_{\bs{t}}^{\pm}\in\BS'$ in Appendix A. 
It's worth noting that $Q_{\bs{t}}^{\pm}$
are linearly independent entire functions whose Wronskian can be evaluated explicitly, \textit{cf}. Lemma \ref{Lemme Wronskien Q t plus moins}: 
\beq
W\pac{Q_{\bs{t}}^+,Q_{\bs{t}}^-}\pa{\la} = \kappa^{-i\la} g^{-N\hbar} \ex{-2N\f{\pi}{\hbar}\la} \cdot
\pl{a=1}{N} \paa{ \f{\hbar}{i\pi} \sinh\f{\pi}{\hbar}\pa{\la-\tau_k} \cdot \mc{H}\pa{\la} }\;,
\label{ecriture Wronskien determinant de Hill}
\enq

It can be shown that the most general solution $q(\la)\in\BS'$ 
to the Baxter equation 
\rf{BAX} may be represented in the form \rf{ansatz}. The 
additional requirement 
that $q(\la)$ should be entire implies 
%
%
%
\begin{equation}\label{q-cond}
Q_{\bs{t}}^+(\de_a)-\zeta\, Q_{\bs{t}}^-(\de_a)\,=\,0\,,\quad\text{for}\quad
a=1,\dots,{N}\, \quad \text{and} \; \text{some} \; \zeta \in \Cx, \;\; \abs{\zeta}=1\;,
\end{equation}
to be supplemented by the condition that $\sum_{k=1}^N\de_k=P$ 
\cite{GutzwillerTodaChainSmallNEigenfunctionsI,
GaudinPasquierTQOperatorTodaChain}.
This formulation of the quantization conditions looks fairly involved. It
may be considered as a highly transcendental system of equations
on the parameters $\tau_k$ determining $\bs{t}(\la)$, in which both 
$Q_{\bs{t}}^\pm(\la)$ and $\de_1,\dots,\de_N$ have to be constructed from 
$\bs{t}(\la)$ by means of 
\rf{Qpmdefinition} and
\rf{definition determinant infini Hill},
\rf{equation factorisation H} respectively.



\subsection{Reformulation in terms of solutions to a nonlinear
integral equation}
\label{Section Generating Baxter from TBA}

One may note that the set of parameters in $\de=(\de_1,\dots,\de_N)$ is just
as big as the set of parameters in 
$\tau=(\tau_1,\dots,\tau_N)$ characterizing the
polynomial $\bs{t}(\la)$ appearing on the left hand side of the Baxter 
equation. The form of the quantization 
conditions \rf{q-cond} suggests that it may be useful to 
formulate these conditions directly in terms of the parameters 
$\de=(\de_1,\dots,\de_N)$ with  $\bs{t}(\la)=\bs{t}(\la|\tau\pa{\de})$ being determined
in terms of $\de$ by inverting the relation $\de=\de(\tau)$. 
A more convenient representation of the quantization conditions
\rf{q-cond} would then be obtained if one was able to construct
the solutions $Q_{\bs{t}}^\pm(\la)$ more 
directly as functions of the parameters
$\de=(\de_1,\dots,\de_N)$. In the
following, for a given polynomial 
$\vth(\la)=\prod_{k=1}^N(\la-\de_k)$ with complex conjugated roots, we will 
construct functions $Q_\de^\pm(\la)$. These will be shown to 
yield solutions the Baxter equation \rf{BAX} via \rf{qfromQ},
with $t_\de(\la)$ being a polynomial
whose coefficients depend on the parameters $\de$. 

The functions $Q_\de^\pm(\la)$ will be build  out of the solutions
$Y_\de(\la)$ to the following NLIE,
\begin{equation}\label{TBA}
\log Y_\de(\la)\,=\,\int_{\BR}d\mu\;K(\la-\mu)\,\ln\left(1+
\frac{\rho^\hbar \,Y_\de(\mu)}{|\vth(\mu-i\hbar/2)|^2}\right)\,,
\end{equation}
where
\begin{equation}\label{kerneldef}
K\pa{\la} = \f{ \hbar } { \pi\pa{\la^2+\hbar^2} }  \;.
\end{equation}

It will be shown in Appendix \ref{existence/uniqueness} that 
the solutions 
$Y_\de(\la)$ to \rf{TBA} 
are unique, and that they exist for all
tuples $\de=(\de_1,\dots,\de_N)$
of zeros of Hill determinants
$\CH(\la)$ constructed from polynomials $\bs{t}(\la)$ whose
zeroes $\tau_k$ satisfy $\abs{\Im\pa{\tau_k}}<\tf{\hbar}
{2}$. The function $Y_\de(\la)$ is meromorphic, 
with its poles accumulating in the direction $\abs{\e{arg}\pa{\la}}=\tf{\pi}{2}$ and such that
$
Y_{\de} \tend 1$ if $\la \tend \infty$
for $\la$ uniformly away from its set of poles.
The properties of $Y_{\de}$ allow one to define two auxiliary functions:
\begin{equation}
\begin{aligned}
\ln v_{\uparrow}\pa{\la}  = - &\Int{\R}{}  \f{\dd \mu}{2i\pi}  
\f{1}{ \la-\mu +i \tf{\hbar}{2}}\left(1+ \f{ \rho^{\hbar} Y_{\de}\pa{\mu} }{ \vth\pa{\mu-i\tf{\hbar}{2}}  \vth\pa{\mu+i\tf{\hbar}{2}}  }\right)\,,
\\
\ln v_{\downarrow}\pa{\la-i\hbar}  =  &\Int{\R}{}  \f{\dd \mu}{2i\pi}  \f{1}{ \la-\mu -i \tf{\hbar}{2}} \left(1+ \f{ \rho^{\hbar} Y_{\de}\pa{\mu} }{ 
\vth\pa{\mu-i\tf{\hbar}{2}}  \vth\pa{\mu+i\tf{\hbar}{2}}  }\right) \;.
\label{definition v up-down section vth}
\end{aligned}
\end{equation}
Out of $v_{\uparrow}\pa{\la}$ and $v_{\downarrow}\pa{\la-i\hbar}$,
we may then construct
\begin{equation}
Q_\de^+(\la)\, =\,    \f{\pa{ \kappa g^N}^{-i\la} \hbar^{i\f{N\la}{\hbar} } \ex{-\f{N\pi}{\hbar}\la} v_{\uparrow}\pa{\la}}
{\pl{k=1}{N} \Ga\pa{1-i\tf{\pa{\la-\de_k}}{\hbar}}  } 
\label{definition Q vth plus/moins} 
\,,\quad
Q_\de^-(\la)\, =\,  \f{ g^{iN\la} \hbar^{-i\f{N\la}{\hbar} }  \ex{-\f{N\pi}{\hbar}\la} v_{\downarrow}\pa{\la-i\hbar} }
{\pl{k=1}{N}  \Ga\pa{1+i\tf{\pa{\la-\de_k}}{\hbar}}  }\,.
\end{equation}
It is shown in Appendix \ref{TBA vs. BAX}
that the functions $Q_{\de}^{\pm}$ are entire 
(\textit{cf}. Lemma \ref{Lemme propriete Q vth}). 
It is also shown there that 
the functions 
$q_\de^\pm(\la)$ defined from $Q_\de^\pm(\la)$ by relations
like \rf{qfromQ} are solutions to the Baxter equation 
\rf{BAX} which have the right asymptotic behavior to be 
contained in $\BS'$.

One may therefore take the construction of $Q_\de^\pm(\la)$ from 
the solutions of the nonlinear integral equation \rf{TBA} 
as a replacement for the construction based on Gutzwillers solutions.
The quantization conditions \rf{q-cond} may now be rewritten
in terms of $Y_\de(\la)$ in the form
\begin{align}\label{ecriture conditions de quantification'}
2\pi n_k & = \f{N \de_k}{\hbar} \ln \hbar  - \de_k \ln \rho + i \ln \zeta
-i \sul{p=1}{N} \ln \frac{\Gamma(1+i\tf{\pa{\de_k-\de_p}}{\hbar})}
{\Gamma(1-i\tf{\pa{\de_k-\de_p}}{\hbar})} \\
& \quad+\Int{\R}{} \f{\dd \tau}{2\pi} \paa{ \f{1}{ \de_k-\tau+i\tf{\hbar}{2}}  +\f{1}{ \de_k-\tau-i\tf{\hbar}{2}}  }
\ln \pa{  1+ \f{ \rho^{\hbar} Y_\de\pa{\tau} }{ \vth \pa{\tau-i\tf{\hbar}{2}}  \vth \pa{\tau+i\tf{\hbar}{2}}  }   }  \;,
\nonumber\end{align}
as is fully demonstrated in Appendix \ref{TBA vs. BAX}.
This form of the quantization condition may be more convenient for
many applications than the ones previously obtained, equations 
\rf{q-cond}.

\subsection{Solutions to the Baxter equation from the
solutions to a NLIE}

In order to understand how the connection between nonlinear integral equations
and the Baxter equation comes about, the key observation
is that the two functions $q_{\de}^{\pm}$ defined above constitute a 
system of two linearly independent solutions of the
Baxter equation \eqref{BAX}. 
This fact can be deduced from 
the so-called quantum Wronskian equation satisfied by $Q_{\de}^{\pm}$:
\begin{align}
Q_{\de}^+\pa{\la}Q_{\de}^-\pa{\la+i\hbar} -Q_{\de}^-\pa{\la}Q_{\de}^+\pa{\la+i\hbar}  =\kappa^{-i\la}  
\bigg(\frac{\hbar \ex{-\f{2\pi\la}{\hbar}}}{ i \pi g^{\hbar}}\bigg)^N 
\pl{k=1}{N} \sinh\f{\pi}{\hbar}\pa{\la-\de_k} \; .
\label{q-Wronski}
\end{align}

The quantum Wronskian equation allows one to show that $Q_\de^\pm(\la)$
 satisfy a Baxter-type equation 
\begin{equation}\label{BAX'}
t_{\de}(\la)Q_\de^\pm(\la)\,=\,i^{-N}g^{N\hbar}Q_\de^\pm(\la+i\hbar) +\kappa^{\hbar}(i)^Ng^{N\hbar}Q_\de^\pm(\la-i\hbar)\,,
\end{equation}
where the polynomial $t_{\de}\pa{\la}$ is defined by
\beq
t_{\de}\pa{\la}= \pa{i\kappa g^{N}}^{\hbar}   
\f{ Q_{\de}^{+}\pa{\la-i\hbar} Q_{\de}^{-}\pa{\la+i\hbar}  - Q_{\de}^{+}\pa{\la+i\hbar} Q_{\de}^{-}\pa{\la-i\hbar}  }
{Q_{\de}^+\pa{\la}Q_{\de}^-\pa{\la+i\hbar} -Q_{\de}^+\pa{\la+i\hbar}Q_{\de}^-\pa{\la}  }  \; .
\label{t-defn}
\enq
On the one hand, the Wronskian relation \rf{q-Wronski} 
 allows one to show
that $t_{\de}(\la)$ and $Q^{\pm}_\de(\la)$ are related by the Baxter equation
\rf{BAX'}. On the other hand,  it also ensures that the 
residues of the possible poles of $t_{\de}$ \rf{t-defn} 
vanish. Then, the polynomiality of $t_{\de}$ is a consequence of the asymptotic behavior of the functions $Q_{\de}^{\pm}$. 
The details of the arguments are found in Appendix 
\ref{TBA vs. BAX}. 

In order to see how the quantum Wronskian relation is connected
to the NLIE \rf{TBA}, let us, starting from 
a solution $Y_{\de}(\la)$ to \rf{TBA}, introduce two
functions $v_{\uparrow}(\la)$ and $v_{\downarrow}(\la)$ 
\textit{via}  \rf{definition v up-down section vth}.
On the one hand, noting that the kernel $K(\la)$ defined in  
\rf{kerneldef} can be written as 
\[
K(\la)\,=\,\frac{1}{2\pi i}\left(\frac{1}{\la-i\hbar}-\frac{1}{\la+i\hbar}\right)
\]
it is easy to see that \rf{TBA} implies
\beq
\ln Y_{\de} \pa{\la} = \ln\pa{v_{\ua}\pa{\la+i\tf{\hbar}{2}}} +  \ln\pa{v_{\da}\pa{\la-3  i\tf{\hbar}{2}}}
\enq
On the other hand, note that
\begin{align}
 & \ln\pac{v_{\ua}\pa{\la-i\f{\hbar}{2} +i0}  } +  \ln\pac{v_{\da}\pa{\la- i\f{\hbar}{2} -i0}}
= \\ 
& \qquad\qquad= \left(\,\int_{\R +i0}-
\int_{ \R - i0}\right)  \f{ \dd \mu}{ 2i\pi }  \f{1} {\la-\mu} 
\left(1+ \f{ \rho^{\hbar} Y_{\de}\pa{\mu} }{ \vth\pa{\mu-i\tf{\hbar}{2}}  \vth\pa{\mu+i\tf{\hbar}{2}}  }\right)\nonumber\\
&\qquad\qquad =1+ \f{ \rho^{\hbar} Y_{\de}\pa{\la} }{ \vth\pa{\la-i\tf{\hbar}{2}}  \vth\pa{\la+i\tf{\hbar}{2}}  } \;.
\nonumber
\end{align}
Thus, using that $v_{\ua/\da}$ are meromorphic on $\Cx$, we are able to continue the obtained relation everywhere on $\Cx$ , leading to the functional 
relation
\beq
v_{\ua}\pa{\la}v_{\da}\pa{\la} = 1+ \f{\rho^{\hbar}}{\vth\pa{\la}\vth\pa{\la+i\hbar} } v_{\ua}\pa{\la+i\hbar}v_{\da}\pa{\la-i\hbar} \;.
\enq
Rewriting this in terms of $Q_\de^\pm$ by means of \rf{definition Q vth plus/moins} yields the quantum Wronskian equation
\rf{q-Wronski}.

At the moment we don't have a direct proof that a solution to \rf{TBA}
exists for all choices of $\vt(\la)$. We are able, however, to prove
that all functions $Y_{\de}(\la)$ that can be constructed from 
Gutzwillers solutions are in fact solutions to \rf{TBA}.
This implies that all functions $Y_{\de}(\la)$ needed for the 
formulation of the quantization conditions 
\rf{ecriture conditions de quantification'} can be obtained 
in this way.


Let us finally note that there is a more direct way (\textit{cf}. 
Proposition \ref{Proposition Polynome de Newton})
to reconstruct 
the Newton polynomials in the zeroes $\paa{\tau_k}$ of $t_{\de}$
from the solution $Y_{\de}$ to \eqref{BAX}:
\begin{align}
\label{Newtonreconstr}
\sul{p=1}{N} \tau^{k}_p
  =\sul{p=1}{N} \de_p^k- k \Int{\R}{} \f{\dd \tau}{2i\pi} & \paa{ \pa{\tau+i\tf{\hbar}{2}}^{k-1}- \pa{\tau-i\tf{\hbar}{2}}^{k-1} }\\[-1.5ex]
& \qquad\qquad\qquad\times
\ln \pa{  1+ \f{ \rho^{\hbar} Y_{\de}\pa{\tau} } { \abs{ \vth\pa{\tau-i\tf{\hbar}{2}} }^2 }   }  \; .
\nonumber
\end{align}
As one may reconstruct the eigenvalues $h_k$ of the conserved quantities 
${\mathbf H}_k$ from the $\sum_{p=1}^{N} \tau^{k}_p$, this essentially
amounts to a reconstruction of the $h_k$.

\subsection{Definition of Yang's potential}

It is interesting to notice that the quantization conditions 
\rf{ecriture conditions de quantification'} characterize the
extrema of a certain function $\mc{W}\pa{{\de}}$ called
Yang's potential in \cite{NekrasovShatashviliConjectureTBADescriptionSpectrumIntModels}.
This Yang's potential is defined as
$\mc{W}\pa{{\de}}=\mc{W}^{\e{inst}}\pa{{\de}} + \mc{W}^{\e{pert}}\pa{{\de}}$, where 
\beq
 \mc{W}^{\e{pert}}\pa{{\de}}  =i \sul{k=1}{N} \f{\de_k^2}{2} \ln\paf{\hbar^{\tf{N}{\hbar}}}{\rho} - \ln \zeta \sul{k=1}{N} \de_k
+\sul{j,k=1}{N} \varpi\pa{\de_k-\de_j}   -2i\pi \sul{k=1}{N} n_k \,,
\enq
where $\varpi^{\prime}\pa{\la}= \ln \Ga\pa{1+i\tf{\la}{\hbar}}$, $n_k$ are some integers paremeterizing the eigenstate, and 
\begin{align}
&\mc{W}^{\e{inst}}\pa{{\de}}=\\
&\quad =-\Int{\R}{} \paa{ \f{\ln Y_\de\pa{\mu}}{2} \ln \pa{ 1+ \f{\rho^{\hbar} Y_\de\pa{\mu}}{ \abs{\vth\pa{\mu+i\tf{\hbar}{2}}}^2 } }
+\e{Li}_2\pa{ \f{-\rho^{\hbar} Y_\de\pa{\mu}}{ \abs{\vth\pa{\mu+i\tf{\hbar}{2}}}^2 } } } \f{\dd \mu}{2i\pi} \;,
\nonumber\end{align}
where 
\beq
\e{Li}_2\pa{z} = \Int{z}{0} \f{ \ln\pa{1-t}}{t} \dd t \;.
\enq
It seems worth emphasizing that, in our treatment, the reformulation 
of the quantization conditions in terms of
Yang's potential was obtained from the solution theory of
the Baxter equation \rf{BAX}. This may be seen as a hint towards
a more direct understanding of the claim in \cite{NekrasovShatashviliConjectureTBADescriptionSpectrumIntModels} that 
the quantization conditions for large classes of quantized
algebraically integrable models can be formulated in this way.
The claim should follow quite generally from the solution theory 
of the Baxter equation.

\appendix

\section{Properties of Gutzwiller's solutions}
\setcounter{equation}{0}

\label{Section NLIE and spectrum}

\subsection{Analytic properties of Gutzwiller's solution}

The explicit construction of a set of two linearly independent entire
solutions $Q_{\bs{t}}^{\pm}$ of \eqref{BAX} with arbitrary monic polynomial $\bs{t}\pa{\la}$
goes back to Gutzwiller \cite{GutzwillerTodaChainSmallNEigenfunctionsI}.
Prior to writing down these two solutions, recall the definition of the  Wronskian of two solutions
$q_1$ and $q_2$ defined by
\beq
W\pac{q_1,q_2}\pa{\la} = q_1\pa{\la} q_2\pa{\la+i\hbar}-q_2\pa{\la} q_1\pa{\la+i\hbar}\,.
\label{definiton du Wronskien}
\enq
It is straightforward to see, using \eqref{BAX}, that  $W\pac{q_1,q_2}$ is $i\hbar$ quasi-periodic:
\beq
W\pac{q_1,q_2}\pa{\la+i\hbar}= \pa{-1}^N\kappa^{\hbar}W\pac{q_1,q_2}\pa{\la} \; .
\label{ecriture quasi-periodicite Wronskien}
\enq

\begin{propn}
\label{Proposition propriete solutions Baxter}
Let $\bs{t}\pa{\la}=\prod_{k=1}^{N}\pa{\la-\tau_k}$ be a monic polynomial of degree $N$ with roots
appearing in complex-conjugate pairs $\paa{\tau_k}=\paa{\ov{\tau}_k}$.
Then, the two functions below are entire solutions to the Baxter equation
\rf{BAX'}, 
\beq
 Q_{\bs{t}}^+(\la)\,=\, \f{ \pa{\kappa g^N}^{-i\la} K_+\pa{\la}  \ex{-N\f{\pi}{\hbar}\la} }{\pl{k=1}{N}\hbar^{-i\frac{\la}{\hbar}} \Ga(1-i(\la-\tau_k)/\hbar)}\,,  \quad 
 Q_{\bs{t}}^-(\la)\,=\,\f{  g^{iN\la} K_-\pa{\la}  \ex{-N\f{\pi}{\hbar}\la}  }{\pl{k=1}{N}\hbar^{i\frac{\la}{\hbar}} \Ga(1+i(\la-\tau_k)/\hbar)}   \,.
\enq
Here $K_{\pm}\pa{\la}$ correspond to the unique meromorphic solutions to 
difference equations %
\beqa
K_+\pa{\la-i\hbar} &=& K_+\pa{\la} -  \f{ \rho^{\hbar} K_+\pa{\la+i\hbar} }{ \bs{t}\pa{\la} \bs{t}\pa{\la+i\hbar} } \; ,
\label{equation de reccurrence K+}\\
K_-\pa{\la+i\hbar} &=& K_-\pa{\la} -  \f{ \rho^{\hbar} K_-\pa{\la-i\hbar} }{ \bs{t}\pa{\la} \bs{t}\pa{\la-i\hbar} } \;.
\label{equation de reccurrence K-}
\eeqa
that go to $1$, 
when $\la\tend \infty$  uniformly away from their set of poles.
The solutions to these recurrence relations
are given explicitly by the determinant formula
\eqref{definition determinant K+} and its complex conjugate. 
These two linearly independent solutions are entire and posses the asymptotic behavior
\beq
\abs{Q_{\bs{t}}^{\pm}\pa{\la}}= \ex{-\f{N\pi}{\hbar}\Re\pa{\la}} \e{O}\big( 
\ex{+\f{N\pi}{2\hbar}\abs{\Re\pa{\la}}}  \abs{\la}^{\f{N}{2\hbar}\pa{ \mp 2\Im\pa{\la}-\hbar } } \big) \quad \e{for} \qquad
\Re\pa{\la}\tend \pm \infty \;,
\enq
and the $\e{O}$ is uniform in every bounded strip of $\Cx$.
\end{propn}

\proof
The only non-trivial part concerns the asymptotic behavior of $K_{\pm}$. It follows as a corollary of Lemma \ref{Lemme AB of K_+ and K_+ prime}. 
\qed
%
%
%
\begin{lem}
\label{Lemme Wronskien Q t plus moins}
The functions $Q_{\bs{t}}^{\pm}$ are linearly independent and their Wronskian
%
%
%
%
%
%
%
%
%
 is expressed in terms of the Hill determinant
\rf{definition determinant infini Hill} by \eqref{ecriture Wronskien determinant de Hill}  .
The latter is closely related to $K_+$ and $K_-$
\beq
\mc{H}\pa{\la} =  K_+\pa{\la} K_-\pa{\la+i\hbar}  -  \rho^{\hbar}  \f{ K_+\pa{\la+i\hbar} K_-\pa{\la } } { \bs{t}\pa{\la} \bs{t}\pa{\la+i\hbar} }  \;.
\label{definition relation de Hill}
\enq
Its zeroes form complex conjugated pairs $\paa{\de_k}=\paa{\ov{\de}_k}$, belong to the fundamental strip $\paa{z \; : \; \abs{\Im\pa{z}}<\tf{\hbar}{2}}$
and fulfill  
$\sum_{p=1}^{N} \tau_p  = \sum_{p=1}^{N} \de_p$.
%
%
%
%
\end{lem}
\proof 

The fact that $Q_{\bs{t}}^{\pm}$ are linearly independent is a consequence of the fact that their Wronskian does not vanish
identically. The explicit expression for this Wronskian follows after some algebra. 

As we have assumed that the set $\paa{\tau_k}$ is self-conjugated, it follows from the determinant representation for $\mc{H}$ that $\ov{\mc{H}\pa{\ov{\la}}}= \mc{H}\pa{\la}$, \textit{ie}, the set $\paa{\de_k}$ is self-conjugated. In its turn, this implies a particular
relation between the set of $\de$'s and $\tau$'s. Namely, computing the $\Re\pa{\la} \to +\infty$ asymptotics of $\mc{H}$ yields
\beq
\sul{p=1}{N} \tau_p  = \sul{p=1}{N} \de_p \; +in\hbar \; ,\quad \e{for} \;\e{some} \quad  n \in \mathbb{N} \; .
\enq
However, as $\sum \tau_k \in \R$ and $\sum \de_k \in \R$, the only possibility is $n=0$ \;. \qed

We are now in position to prove the

\begin{lem}
\label{lemme forme generale solutions TQ}
Let $q$ be any meromorphic solution to \eqref{BAX}. Then, there exists two meromorphic $i\hbar$-periodic functions $P_{\pm}\pa{\la}$  such that
\beq
q\pa{\la}=P_+\pa{\la} Q_{\bs{t}}^+\pa{\la} + P_-\pa{\la}  Q_{\bs{t}}^-\pa{\la}  \;.
\label{forme generale solutions equation TQ}
\enq
\end{lem}

\proof

Let $q$ be any meromorphic solution to Baxter's T-Q equation \eqref{BAX}. Then consider
\beq
\wt{q}\pa{\la}=q\pa{\la}-\f{ W\pac{q,Q_{\bs{t}}^{-}}\pa{\la}  }{ W\pac{Q_{\bs{t}}^+,Q_{\bs{t}}^-}\pa{\la}  } \cdot Q_{\bs{t}}^+\pa{\la}
+ \f{ W\pac{q,Q_{\bs{t}}^+}\pa{\la}  }{ W\pac{Q_{\bs{t}}^+,Q_{\bs{t}}^-}\pa{\la}  } \cdot Q_{\bs{t}}^-\pa{\la} \; .
\enq
The ratio of two Wronskian being  $i\hbar$ periodic, one gets that, by construction
\beq
W\pac{\wt{q},Q_{\bs{t}}^+}\pa{\la}=W\pac{\wt{q},Q_{\bs{t}}^-}\pa{\la}=0 \;.
\enq
This leads to the system of equations for $\wt{q}\pa{\la}$:
\beq
 \pa{\ba{cc} Q_{\bs{t}}^+\pa{\la} &  Q_{\bs{t}}^+\pa{\la+i\hbar} \\   Q_{\bs{t}}^-\pa{\la} &  Q_{\bs{t}}^-\pa{\la+i\hbar}      \ea }
\pa{\ba{cc} -\wt{q}\pa{\la+i\hbar} \\ \wt{q}\pa{\la} \ea } = 0
\label{system eqns pour q tilde}
\enq
Given any fixed $\la$, there exist non-trivial solutions to \eqref{system eqns pour q tilde} if only if the determinant of the matrix defining the system
vanishes, \textit{ie} $W\pac{Q_{\bs{t}}^+,Q_{\bs{t}}^-}\pa{\la}=0$. However, it follows from \eqref{ecriture Wronskien determinant de Hill} that
$W\pac{Q_{\bs{t}}^+,Q_{\bs{t}}^-}\pa{\la}$ is an entire function that is non-identically zero. Therefore, it can only vanish at isolated points.
Hence, we get that $\wt{q}\pa{\la} \not= 0$ only at an at most countable set. As $\wt{q}\pa{\la}$ is meromorphic  on $\Cx$, $\wt{q}=0$. \qed

We now provide a rough  characterization of the set of zeroes of $Q_{\bs{t}}^{\pm}$.
As the $\Ga$ function has no zeroes on $\Cx$, the only zeroes of $Q_{\bs{t}}^{\pm}$ are those of $K_{\pm}\pa{\la}$.

\begin{propn}
\label{Proposition zeroes K pm}
Assume that $\abs{\Im\pa{\tau_k}}< \tf{\hbar}{2}$ and that the set $\paa{\tau_k}$ is invariant under complex conjugation.
Then, the set of zeroes of $K_+\pa{\la-i\tf{\hbar}{2}}$ belongs to the half-plane
$\paa{z \in \Cx \; : \; \Im\pa{z}< -\tf{\hbar}{2}}$ and
\beq
 \f{  \abs{K_+\pa{\la-i\tf{\hbar}{2}}}^2 }{ \mc{H}\pa{\la-i\tf{\hbar}{2}} } 
>1\,, \qquad  \e{for} \quad \la \in \R \;.
\enq
A similar statement holds for $K_-\pa{\la+i\tf{\hbar}{2}}$, namely the set of zeroes of
$K_-\pa{\la+i\tf{\hbar}{2}}$ lies in the half-plane $\paa{z \in \Cx \; : \; \Im\pa{z}> +\tf{\hbar}{2}}$ and
$K_-\pa{\la+i\tf{\hbar}{2}}$  does not vanish on $\R$.
\end{propn}

\proof

It follows from the determinant representations that $K_+\pa{\la-i\tf{\hbar}{2}}$ has poles at $\tau_k-i\pa{2n+1}\tf{\hbar}{2}$, $n\in \mathbb{N}$, in 
particular they all belong to the half-plane $\paa{z \in \Cx \; : \; \Im\pa{z}< -\tf{\hbar}{2}}$. Also, since the zeroes and poles of the Hill determinant
are self-conjugated,
\beq
\mc{H}\pa{\la- i\tf{\hbar}{2}}  = \pl{k=1}{N} \f{ \cosh\f{\pi}{\hbar}\pa{\la-\de_k}  }{  \cosh\f{\pi}{\hbar}\pa{\la-\tau_k}  }  > 0
\quad , \quad  \forall \la \; \in \, \R \:.
\label{equation positivite H sur R}
\enq
%
%
%
%
%
%
%
%
%
%
%
%
%
As the set $\paa{\tau_k}$ is self-conjugate, it is easy to see that
\beq
\ov{K_+\pa{\la}}=K_-\pa{\ov{\la}} \qquad \e{and} \qquad
\bs{t}\pa{\la-i\tf{\hbar}{2}} \bs{t}\pa{\la+i\tf{\hbar}{2}} =\abs{\bs{t}\pa{\la-i\tf{\hbar}{2}} }^2 \; ,
\enq
This allows us to rewrite \rf{definition relation de Hill} in the form
\beq
\f{ \abs{K_+\pa{\la-i\tf{\hbar}{2}}}^2  } { \mc{H}\pa{\la-i\tf{\hbar}{2}} } = 1 +
 \f{ \rho^{\hbar} }{ \mc{H}\pa{\la-i\tf{\hbar}{2}} }
  \abs{  \f{K_+\pa{\la+i\tf{\hbar}{2}}  } {\bs{t}\pa{\la-i\tf{\hbar}{2}}  }  }^2\;.
\label{equation pour K+2 sur H}
\enq
It follows from \eqref{equation positivite H sur R} and \eqref{equation pour K+2 sur H} that there exists a $c>0$
such that $\abs{K_+\pa{\la-i\tf{\hbar}{2}}} >c$ for $\la \in \R$. Hence, $K_+\pa{\la-i\tf{\hbar}{2}}$
has no zeroes on $\R$. As $K_+\pa{\la-i\tf{\hbar}{2}}$ has manifestly no poles in the half-plane $\paa{z \; : \; \Im\pa{z} > -\tf{\hbar}{2} }$, one has that
\beq
f\pa{\rho} = \hspace{-2mm} \Int{ \R-i\tf{\hbar}{2} }{ } \hspace{-2mm} \f{\dd \tau}{2i\pi}  \f{ K_+^{\prime}\pa{\tau} }{ K_+\pa{\tau} } =
\# \paa{ z \in \Cx \; : \; \Im\pa{z} >-\f{\hbar}{2} \quad \e{and} \quad  K_+\pa{z}=0  } \;.
\label{definition fonction nombre zeroes K+}
\enq

Here, we remind that $\rho$ is the deformation parameter appearing in \eqref{definition determinant K+}.
We also specify that the function $f\pa{\rho}$ is well defined as $\abs{K_+}_{\mid\R-i\tf{\hbar}{2}}>c$ and the ratio $\tf{K^{\prime}_+}{K_+}$
decays at least as $\la^{-\pa{2N+1}}$ at infinity, uniformly in $\rho$, \textit{cf} lemma \ref{Lemme AB of K_+ and K_+ prime}. Thus, applying the dominated
convergence theorem we obtain that $f\pa{\rho}$ is continuous in $\rho$. As it is integer valued, it is constant.
The value of this constant is fixed from  $f\pa{0}=0$ (as then $K_+=1$). This shows that  $K_+\pa{\la}$
has all of its zeros lying below the line $\R-i\tf{\hbar}{2}$. \qed


\subsection{Bounds for $K_+$}\label{Appendice AB of K_+ and K_+ prime}

\begin{lem}
\label{Lemme AB of K_+ and K_+ prime}
Let  $\paa{\tau_k}=\paa{\ov{\tau}_k}$ and $\abs{\Im \pa{\tau_k}}<\tf{\hbar}{2}$, then $K_{\pm}\rightarrow 1$ for $\la \rightarrow \infty$ uniformly away from its set of poles and 
\beq
\f{K_{\pm}^{\prime}}{K_{\pm}}\pa{\la}= \e{O}\pa{ \la^{-\pa{2N+1}}} \; ,
\enq
where the $\e{O}$ is uniform as long as $\rho$ belongs to some fixed compact subset of $\Cx$. 

\end{lem}

\proof

As the T-Q equations can be solved explicitly when $N=1$ in terms of Bessel functions, 
it is enough to consider the case $N\geq 2$. We focus on $K_+$ as behavior of $K_-$ follows by complex conjugation.

$K_+$ admits the discrete Fredholm series representation:
\beq
K_+\pa{\la} = 1+  \sul{n\geq 1}{} \f{1}{n!} \sul{h_1,\dots, h_n \in \mathbb{N}}{} \ddet{n}{M_{h_ah_b}\pa{\la}} \; ,
\enq
where we have
\beq
M_{ab}\pa{\la} = \f{\de_{a,b+1} }{\bs{t}\pa{\la-i a \hbar}} + \f{\de_{a,b-1}  \rho^{\hbar} }{\bs{t}\pa{\la-i a \hbar}} \;.
\enq
Then, by Haddamard's inequality
\bem
\abs{\sul{h_1,\dots, h_n \in \mathbb{N}}{} \ddet{n}{M_{h_ah_b}\pa{\la}} } \leq \sul{h_1,\dots, h_n \in \mathbb{N}}{}
\pl{a=1}{n} \paa{ \sul{b=1}{n} \abs{M_{h_a h_b}}^2}^{\f{1}{2}} \\
\leq \pl{a=1}{n} \paa{ \sul{a,b=1}{n} \abs{M_{h_a h_b}}} \leq u^n\pa{\la} \pa{1+\abs{\rho}^{\hbar}}^n \;.
\end{multline}
There we have set
\beq
u\pa{\la}=\sul{k=1}{+\infty} \abs{\bs{t}\pa{\la-ik\hbar}}^{-1}  \leq \paf{2}{\hbar}^{N} \sul{k=1}{+\infty} k^{-N} 
\qquad \text{and} \quad u\pa{\la}=\e{O}\pa{\la^{-\f{1}{2}}}\; .
\enq
Hence, we get that 
\beq
\abs{K_+\pa{\la}-1} \leq \ex{u\pa{\la} \pa{1+\abs{\rho}^{\hbar}}} -1 \;,
\enq
and thus $K_+ \rightarrow 1$ for $\la \rightarrow \infty$ uniformly away from its set of poles. 
It remains to provide the stronger estimates for its decay at infinity. 

Termwise differentiation of the series leads to 
\[
\abs{K^{\prime}_+\pa{\la}} \leq  \sul{n\geq 1}{} \f{n}{n!} \pa{u\pa{\la} 
\big(1+\abs{\rho}^{\hbar}\big)^{n-1}} \wt{u}\pa{\la} 
\big(1+\abs{\rho}^{\hbar}\big) \leq 
\wt{u}\pa{\la} \pa{1+\rho^{\hbar}} \ex{u\pa{\la} \pa{1+\abs{\rho}^{\hbar}}} \;,
\]
where $\wt{u}\pa{\la}=\sum_{n\geq 1} \abs{ \tf{\bs{t}^{\prime}} { \bs{t}^{2}} \pa{\la-in\hbar} }  = \e{O}\pa{\abs{\la}^{-1}}$. 
One then takes the derivative of the Hill's determinant relation 
\rf{definition relation de Hill} at $\la-i\tf{\hbar}{2}$. This leads to 
\beq
K_+^{\prime}\pa{\la} = \mc{H}^{\prime}\pa{\la} + \Dp{\la} \paa{ \rho^{\hbar} \f{K_+\pa{\la+i\hbar}K_-\pa{\la} } 
{\bs{t}\pa{\la}\bs{t}\pa{\la+i\hbar}K_-\pa{\la+i\hbar}} } \;.
\enq
The uniform estimates in $\rho$ that we have established combined with the fact that  $\mc{H}^{\prime}\pa{\la}=\e{O}\pa{\la^{-\infty}}$
uniformly in $\rho$ lead to the desired form of the estimates, with a $\e{O}$ that is uniform as long as $\rho$ belongs to 
some compact subset of $\Cx$. \qed 

\section{Existence and uniqueness of solutions to the  
nonlinear integral equation}
\setcounter{equation}{0}

\label{existence/uniqueness}

In this Appendix we prove the existence and uniqueness of solutions to the TBA non-linear integral equation \rf{BAX'}. 
Let $\vt\pa{\la}=\prod_{k=1}^{N}\pa{\la-\de_k}$ have its zeroes
given by the $N$ zeroes of the Hill determinant 
built out of $\bs{t}(\la)$. Then set 
\beq
Y_{\bs{t}}\pa{\la} =  \f{ K_+\pa{\la+i \tf{\hbar}{2}}K_-\pa{\la-i \tf{\hbar}{2}} }{ \mc{H}\pa{\la-i \tf{\hbar}{2}} }\cdot
\f{ \vt\pa{\la-i \tf{\hbar}{2}} \vt\pa{\la+i \tf{\hbar}{2}} }
{ \bs{t}\pa{\la - i \tf{\hbar}{2}} \bs{t}\pa{\la+i \tf{\hbar}{2}} }  \;.
\label{definition Y pour NLIE}
\enq
It is a straightforward consequence of Lemma \ref{Lemme AB of K_+ and K_+ prime} that $Y_{\bs{t}}$ is a meromorphic function
whose poles accumulate in the direction $\abs{\arg\pa{\la}}=\tf{\pi}{2}$ and such that $Y_{\bs{t}} \tend 1$ for $\la \rightarrow \infty$
uniformly aways from the set of its poles.

\begin{propn}
\label{Proposition Unicite et existence solutions}
The function  
$\ln Y_{\bs{t}}$ defined in 
\rf{definition Y pour NLIE}
is continuous, positive and bounded on $\R$. 
It is the unique solution in this class to the 
non-linear integral equation \rf{TBA}.
\end{propn}

\begin{proof}
We first prove the uniqueness of solutions. Let $\norm{\cdot}_{\infty}$ stand for the sup norm on bounded and continuous
functions on $\R$. We set $\mc{F}=\paa{ f \in \msc{C}^0\pa{\R} \; : \; f \geq 0 \; \quad \e{and} \quad \norm{f}_{\infty}<+\infty  }$.
Then we define the operator $L$ on $\mc{F}$ by
\beq
L\pac{f}\pa{\la} = \Int{\R}{} \!\! \dd \mu \;  K\pa{\la-\mu}
\ln\pa{ 1+ \f{  \rho^{\hbar} \ex{f\pa{\mu}}  }{  \abs{\vt\pa{\la-i \tf{\hbar}{2}}  }^2   } } \;.
\enq

The mapping $L$ stabilizes $\mc{F}$. Indeed,
\beq
\abs{L\pac{f}\pa{\la}} \leq \Int{\R}{} \!\! \dd \mu \;  K\pa{\la-\mu}
\f{  \rho^{\hbar} \ex{f\pa{\mu}}  }{  \abs{\vt \pa{ \la-i \tf{\hbar}{2} }  }^2   } \leq 
\ln\pa{ 1+ \rho^{\hbar} \ex{\norm{f}_{\infty}} J^{-1}  } \; ,
\enq
where $J=\inf_{\la \in \R}  \abs{ \vt  \pa{\la-i\tf{\hbar}{2}} }^2 >0$, due to $\abs{\Im\pa{\de_k}}<\tf{\hbar}{2}$.

Any solution to the NLIE appears as a fixed point of $L$ in $\mc{F}$. We shall now prove that
$L$ can have at most one fixed point. This settles the question of uniqueness of solutions to \eqref{TBA}.
This part goes as in \cite{FringKorffSchultzSomeAnalysisOFNLIEandMore}.
Let $f, g \in \mc{F}$, then
\bem
\abs{L\pac{f}-L\pac{g}}\pa{\la} = \abs{ \Int{0}{1} \!\!\dd t \Int{\R}{} \!\! \dd \tau K\pa{\la-\tau}
\f{ \rho^{\hbar} \ex{g\pa{\tau}+t\pa{f-g}\pa{\tau}}   }
{ \abs{ \vt\pa{\tau+i\tf{\hbar}{2}} }^2 + \rho^{\hbar} \ex{g\pa{\tau}+t\pa{f-g}\pa{\tau}}     }   \pa{f-g}\pa{\tau}
} \\
\leq  \f{ \rho^{\hbar} \ex{ \max\pa{ \norm{g}_{\infty} , \norm{f}_{\infty} } }  }
{  J +  \rho^{\hbar} \ex{ \max\pa{ \norm{g}_{\infty} , \norm{f}_{\infty} } }  }  \norm{f-g}_{\infty} < \norm{f-g}_{\infty}\;.
\label{equation de L comme contraction}
\end{multline}
Hence, $L$ admits a unique fixed point.


A direct proof of existence of the solutions to \rf{TBA} is 
possible if  $\tf{\rho^{\hbar}}{J}<1$. In this case 
it is easily seen that  $L\pac{f}\pa{\la}$ 
is a bounded mapping in the sense that
it stabilizes all balls in $\mc{F}$ of radius $R\geq -\ln\pa{1-\tf{\rho^{\hbar}}{J}}$.
In such a case, \eqref{equation de L comme contraction} implies that $L\pac{f}$ is a contractive map on a Banach space.
It thus admits a unique fixed point.

However, it is always possible 
to construct a solution to \eqref{TBA} 
in terms of the the half-infinite 
determinants $K_{\pm}$. Recall that $\paa{\tau_k}$ and 
hence $\paa{\de_k}$ are invariant under complex conjugation.
Let
\beqa
v_{\uparrow}\pa{\la} = K_+\pa{\la} \pl{k=1}{N} 
\frac{\Gamma(1-i\tf{\pa{\la-\de_k}}{\hbar})}{\Gamma(1-i\tf{\pa{\la-\tau_k}}{\hbar})}
\label{definition v up}\\
v_{\downarrow}\pa{\la} = K_-\pa{\la+i\hbar} \pl{k=1}{N} \frac{\Gamma(i\tf{\pa{\la-\de_k}}{\hbar}) }{\Gamma(i\tf{\pa{\la-\tau_k}}{\hbar} ) } \; .
\label{definition v down}
\eeqa

It follows from Proposition \ref{Proposition zeroes K pm} that $v_{\uparrow/\downarrow}\pa{\la-i\tf{\hbar}{2}}$ are holomorphic
and non-vanishing in $\ov{\mathbb{H}}_{+/-}$. Moreover, as $\sum \de_k=\sum \tau_k$, we get that $v_{\uparrow/\downarrow}\pa{\la-i\tf{\hbar}{2}}=
1+\e{O}\pa{\la^{-1}}$ in their respective domains of holomorphy.
Also, due to the Hill determinant identity \eqref{definition relation de Hill}
\beq
\f{ \abs{K_+\pa{\la-i\tf{\hbar}{2}}}^2  }{ \mc{H}\pa{\la}}=
v_{\uparrow}\pa{\la-i\tf{\hbar}{2}} v_{\downarrow}\pa{\la-i\tf{\hbar}{2}}
=  1+ \f{ \rho^{\hbar} Y_{\bs{t}}\pa{\la} }{ \abs{ \vt \pa{\la-i\tf{\hbar}{2}} }^2 }
\enq
We agree upon choosing a determination of $v_{\uparrow/\downarrow}\pa{\la-i\tf{\hbar}{2}}$ such that
\[
\ln v_{\uparrow/\downarrow}\pa{\la-i\tf{\hbar}{2}} \limit{\la}{+\infty} 0
\Rightarrow \ln \pac{v_{\uparrow}v_{\downarrow}}\pa{\la-i\tf{\hbar}{2}} =
\ln v_{\uparrow}\pa{\la-i\tf{\hbar}{2}}  +   \ln v_{\downarrow}\pa{\la-i\tf{\hbar}{2}} \; .
\]

Thus, for $\la \in \R$,  by computing the residues in the upper or lower half-plane and using the
decay properties of the integrand at infinity, one sees that 
$v_\uparrow$ and $v_\downarrow$ are recovered from $Y_{\bs{t}}(\la)$ via
the integral representations \rf{definition v up-down section vth}.
%
%
%
%
%
%
%
%
%
%
%
%
%
%
%
%
Hence, with the same choice of branches of logarithm as before (the one that goes to $0$ when $\Re\pa{\la}$ goes to $+\infty$) we get, on the one hand, 
that
\beq
\ln\pac{  v_{\uparrow}\pa{\la+i\tf{\hbar}{2}} v_{\downarrow}\pa{\la-3i\tf{\hbar}{2}}  } =
\Int{\R}{} \dd \tau K\pa{\la-\tau} \ln \pa{  1+ \f{ \rho^{\hbar} Y_{\bs{t}}\pa{\tau} }{ \abs{ \vt \pa{\tau-i\tf{\hbar}{2}} }^2 }   }  \; .
\enq

On the other hand, it is straightforward to check that
$
v_{\uparrow}\pa{\la+i\tf{\hbar}{2}} v_{\downarrow}\pa{\la-3i\tf{\hbar}{2}} = Y\pa{\la} .
$
This proves the existence of the relevant set of solutions to \eqref{TBA}. 
\end{proof}

\section{Baxter Equation and quantization conditions from TBA}
\label{TBA vs. BAX}
\setcounter{equation}{0}

We first prove basic properties of the functions $Q_{\de}^{\pm}$.
Then we derive the T-Q equation generated by $Q_{\de}^{\pm}$
and finally obtain the quantization conditons. 

\subsection{Analytic properties of $Q_{\de}^{\pm}$}
\begin{lem}
\label{Lemme propriete Q vth}
The functions $Q_{\de}^{\pm}$ defined in
\eqref{definition Q vth plus/moins}  are entire and have the asymptotic behavior
\beq
\abs{ Q_{\de}^{\pm} }= \ex{-\f{N\pi\la}{\hbar} }  \, \cdot \, \e{O}\big( 
\ex{\f{N\pi}{2\hbar}\abs{\Re\pa{\la}}}  
\abs{\la}^{\f{N}{2\hbar}\pa{ \pm 2\Im\pa{\la}-\hbar } } \big)
\qquad  \Re\pa{\la} \tend \pm \infty \;
\label{equation AB Qvth pm}
\enq
where the $\e{O}$ symbol is uniform in $\paa{z \; : \; \abs{\Im\pa{z}}\leq \tf{\hbar}{2}}$.
\end{lem}
\proof

It is readily seen from the asymptotic behavior in the strip $\paa{z \; : \; \abs{\Im\pa{z}} < \hbar}$  of the solution $Y_{\de}$ to \eqref{TBA},
that $v_{\ua}\pa{\la} \tend 1$ and  $v_{\ua}\pa{\la-i\hbar} \tend 1$  when $\Re\pa{\la}\tend \pm \infty$ in the strip $\paa{z \; : \; \abs{\Im\pa{z}}\leq
\tf{\hbar}{2}}$. Then a straightforward computation leads to \eqref{equation AB Qvth pm}.
We assume that all the $\de$'s are distinct and, if necessary, take the limit of coinciding $\de$'s at the end of the calculation. 

It remains to prove that $Q_{\de}^{\pm}$ are entire. 
For this, we show that the products of $\Ga$-functions cancel the poles
of $v_{\ua/\da}$.
We need to construct a meromorphic continuation to $\Cx$ of $v_{\ua/\da}$  starting from the strip
$\mc{B}^{\pa{1}}$, with $\mc{B}^{\pa{n}}=\paa{z \; : \;  \abs{\Im\pa{z}}< n \hbar  }$. It follows from the very form of the NLIE \eqref{TBA}
that the solution $Y_{\de}$ is holomorphic in $\mc{B}^{\pa{1}}$.
Let us introduce the notation
\begin{equation}
V_\de(\mu)=1+
\frac{\rho^\hbar \,Y_\de(\mu)}{|\vth(\mu-i\hbar/2)|^2}\,.
\end{equation}
The only singularities of 
$V^{\prime}_{\de}\pa{\mu}/V_{\de}\pa{\mu}$ in $\mc{B}^{\pa{1}}$ correspond to the zeroes of $V_{\de}$ and to its 
poles. The latter are located at $\mu=\de_k \pm i\tf{\hbar}{2}$, $k=1,\dots, N$. Hence, as $V_{\de}^{\prime}\pa{\mu}/V_{\de}\pa{\mu}$ is decaying sufficiently fast at infinity,
one gets 
%
%
%
%
%
%
%
%
%
%
%
%
%
%
%
%
%
\begin{align}\label{equation poleet zeroY strip n}
\f{Y_{\de}^{\prime}}{Y_{\de}}\pa{\tau} & = \sul{ \substack{ z \in \mc{B}^{\pa{n}}_{\ua} \\ V_{\de}\pa{z}=0}  }{} \f{ n_z }{ \la-z-i\hbar  }
+ \sul{ \substack{ z \in \mc{B}^{\pa{n}}_{\da} \\ V_{\de}\pa{z}=0}  }{} \f{ n_z }{ \la-z+i\hbar  }
\\ & \quad-\sul{p=1}{n}\sul{k=1}{N} \f{ 1 }{\la-\de_k-\tf{i\pa{2p+1}\hbar}{2}} 
-\sul{p=1}{n}\sul{k=1}{N}  \f{ 1 }{\la-\de_k+\tf{i\pa{2p+1}\hbar}{2}}
\nonumber\\
& \quad + \Int{ \R+in\hbar- i0^+  }{} \f{\dd \mu}{2i\pi} \f{ \tf{V_{\de}^{\prime}\pa{\mu}}{V_{\de}\pa{\mu} }} { \la-\mu-i\hbar }
-\Int{ \R-in\hbar + i0^+  }{} \f{\dd \mu}{2i\pi} \f{\tf{V_{\de}^{\prime}\pa{\mu}}{V_{\de}\pa{\mu}}} { \la-\mu+i\hbar }  \;.
%
%
\nonumber
\end{align}
Above, we have denoted by $n_z$ the multiplicity of a zero $z$ of $V_{\de}$ and $\mc{B}^{\pa{n}}_{\ua}$, resp.  $\mc{B}^{\pa{n}}_{\da}$, stands for
$\mc{B}^{\pa{n}} \cap \mathbb{H}_+$, resp.  
$\mc{B}^{\pa{n}} \cap \mathbb{H}_-$.
It thus follows that $V_{\de}$ has simple poles at 
$\de_k+i\frac{\hbar}{2}\pa{2n+1}$, $n\in \mathbb{Z}$.
Exactly the same reasoning as before shows that $v_{\ua}\pa{\la-i\hbar}$ has its only simple poles
at $\de_k-in\hbar$, $k=1,\dots,N$ and $n\in \mathbb{N}^*$. Similarly,
$v_{\da}\pa{\la}$ has its only simple poles
at $\de_k+in\hbar$, $k=1,\dots,N$ and $n\in \mathbb{N}$.
Therefore, these poles are canceled out by the zeroes of the $\Ga$-functions and  $Q_{\de}^{\pm}$ are both entire. \qed

\subsection{Baxter equation}\label{BAXsubsec}

\begin{propn}
\label{Proposition eqn TQ vth}
The polynomial $t_{\de}\pa{\la}$ given in \eqref{t-defn}  is a monic real valued polynomial of degree $N$.
It is such that the functions $Q_{\de}^{\pm}$ solve the Baxter equation
\begin{equation}\label{BAX''}
t_{\de}(\la)\,Q_\de^\pm(\la)\,=\,i^Ng^{N\hbar}\,Q_\de^\pm(\la+i\hbar)
+\kappa^{\hbar}(-i)^Ng^{N\hbar}\,Q_{\de}^\pm(\la-i\hbar)\,,
\end{equation}
Finally, given any monic real valued polynomial of degree $N$ 
with roots in the strip $\paa{z \; :\; \abs{\Im\pa{z}}<\tf{\hbar}{2}}$, 
it is always possible to find a complex-conjugation invariant set of parameters $\paa{\de_k}$
such that $t_{\de}$ equals to this polynomial.
\end{propn}

\proof
We first prove that  $Q_{\de}^{+}$ satisfies \eqref{BAX''} with $t_\de$ 
being given by \eqref{t-defn}.
\begin{align}
\label{tQcalculattion}
&\kappa^{-i\la} \ex{ -\f{2\pi N}{\hbar}\la } \pl{p=1}{N}\paa{\f{\hbar}{\pi} \sinh\f{\pi}{\hbar}\pa{\la-\de_p} }  \cdot
t_{\de}\pa{\la}Q_{\de}^{+}\pa{\la} \\
&= \pa{\kappa g^{2N}}^{\hbar}  Q_{\de}^{+}\pa{\la-i\hbar}  
\paa{\kappa^{-i\la}  \paf{\hbar \ex{-\f{2\pi\la}{\hbar}}  }{ i \pi g^{\hbar}}^N \pl{k=1}{N} \sinh\f{\pi}{\hbar}\pa{\la-\de_k}  + Q_{\de}^{+}\pa{\la+i\hbar} Q_{\de}^{-}\pa{\la}   } \nonumber\\
&\;\;-\pa{\kappa g^{2N}}^{\hbar}  Q_{\de}^{+}\pa{\la+i\hbar} 
\paa{Q_{\de}^{+}\pa{\la-i\hbar} Q_{\de}^{-}\pa{\la}  - \kappa^{-i\la-\hbar}  \paf{ i \hbar \ex{-\f{2\pi\la}{\hbar}}  }{ \pi g^{\hbar}}^N \pl{k=1}{N} \sinh\f{\pi}{\hbar}\pa{\la-\de_k}  } \nonumber\\
&=\kappa^{-i\la} \ex{ -\f{2\pi N}{\hbar}\la } \pl{p=1}{N}\paa{\f{\hbar}{\pi} \sinh\f{\pi}{\hbar}\pa{\la-\de_p} }
\paa{g^{N\hbar} i^N Q_{\de}^+\pa{\la+i\hbar} + \kappa^{\hbar} g^{N\hbar} \pa{-i}^N Q_{\de}^+\pa{\la-i\hbar} }.
\nonumber\end{align}
In the intermediate steps, we have used the definition of $t_{\de}$ and the
q-Wronskian equation \rf{q-Wronski}.

Now we show that $t_{\de}$, as defined in \eqref{t-defn}, is indeed a monic real valued polynomial of degree $N$.
We first assume that the $\de_k$'s are pairwise distinct. The case when several $\de_k$'s coincide follows by taking the limit in the final formulae.
As $Q_{\de}^{\pm}$ are both entire, we get that the only potential poles of $t_{\de}\pa{\la}$ are located at $\la=\de_k+in\hbar$, $n\in\mathbb{Z}$.
The set of zeroes of $v_{\ua/\da}$ differs necessarily from its set of poles (as follows readily from \eqref{equation poleet zeroY strip n}
and similar representations for $v_{\ua/\da}$). Hence, $Q_{\de}^{\pm}\pa{\de_k+in \hbar}\not=0$, for $k=1,\dots,N$
and $n \in \mathbb{Z}$. Therefore, it follows from the Wronskian relation \eqref{q-Wronski} , that
\beq
\f{Q_{\de}^{+}\pa{\de_k}}{Q_{\de}^{-}\pa{\de_k}}= \f{Q_{\de}^{+}\pa{\de_k+in\hbar}}{Q_{\de}^{-}\pa{\de_k+in\hbar}} \quad \e{for} \qquad k\in\intn{1}{N} \;\; \e{and}
\quad n\in \mathbb{N}\; .
\enq
This implies that the possible poles of the expression in \rf{t-defn}
get canceled.
%
%
%
%
 %
%
%
%
%
It follows that $t_{\de}$ is entire.
It remains to control its asymptotic behavior.  
We may express $t_{\de}$ in terms of $v_{\ua/\da}$, 
\beq
t_{\de}\pa{\la}= v_{\ua}\pa{\la-i\hbar} v_{\da}\pa{\la} \pl{a=1}{N} \pa{\la-\de_a} -
\f{\pa{\kappa g^{2N} }^{2\hbar} v_{\ua}\pa{\la+i\hbar} v_{\da}\pa{\la-2i\hbar} }
{\pl{a=1}{N} \pa{\la-\de_a}  \pa{\hbar^2+\pa{\la-\de_a}^2} } \;.
\enq
Due to the asymptotic behvavior of $Y_{\de}$ at $\infty$, one can deform the integration contour in the definition $v_{\ua/\da}$
so as to obtain its asymptotic behavior in the whole plane $\la\rightarrow \infty$, for $\la$ uniformly away from the set of poles of 
$v_{\ua/\da}$. 

As we have that $v_{\ua/\da} \tend 1$ when $\la\tend \infty$, we get that $t_{\de} \simeq \la^N$ when $\la\tend \infty$.
Hence, $t_{\de}$ is a monic polynomial of degree $N$. It is real valued for $\la\in \R$ as for such $\la$'s, $V_{\th}\pa{\la} \in\R$,
what implies that $v_{\ua}\pa{\la-i\hbar}=v_{\da}\pa{\la}$, \textit{ie} $\ov{t_{\de}\pa{\la}}= t_{\de}\pa{\ov{\la}}$. 

The fact that,  in this way,
one is able to generate any monic polynomial with roots in the 
strip $\paa{z \; : \; \abs{\Im\pa{z}}< \tf{\hbar}{2}}$ follows
form the uniqueness of solutions to the TBA-NLIE and the 
construction of the function $v_{\ua/\da}$ in terms of 
determinants, as  
given in  \eqref{definition v up}-\eqref{definition v down}. \qed


\subsection{The quantization conditions}\label{SSec:q-cond}

We now prove that the quantization conditions for the model (conditions
on the zeroes of the polynomial $t_{\de}\pa{\la}$ for \eqref{BAX} to have entire solutions with a prescribed decay as given in point (ii))
can be written down in a TBA-like form. Moreover, 
as opposed to the Gutzwiller
form of the quantization 
conditions, the ones that will follow only involve one set of parameters. Namely, the zeroes $\paa{\de_k}$ of the Hill determinant associated with
$t_{\de}\pa{\la}$ given in \eqref{t-defn}. We show that under certain reasonable assumptions, it is possible to
reconstruct the Newton polynomials in the zeroes of $t_{\de}$ and hence the spectrum of the model.
This proves the Nekrasov-Shatashvili conjecture \cite{NekrasovShatashviliConjectureTBADescriptionSpectrumIntModels}. We first reconstruct the zeroes of 
$t_{\de}$.

\begin{propn}
\label{Proposition Polynome de Newton}

Let $t_{\de}\pa{\la}=\prod_{p=1}^{N} \pa{\la-\tau_a}$ be a polynomial whose zeroes $\tau_k$ lie in the strip
$\paa{ z \in \Cx \; : \; \abs{\Im\pa{z}} < \tf{\hbar}{2} }$. If $\paa{\de_k}$ is the associated set of zeroes of the Hill determinant
and $Y_{\de}$ the unique solution to the NLIE \eqref{TBA}, then the Newton polynomials  $\mc{E}_k=\sum_{p=1}^{N} \tau_p^k$
in the zeroes of $t_{\de}$ are reconstructed by means of formula
\rf{Newtonreconstr} above.
%
%
%
%
%
%
%
%
%
The convergence of these integrals is part of the conclusion.
\end{propn}

\proof

Due to the uniqueness of solutions to the NLIE \eqref{TBA}, one has that the solution $Y_{\de}$ can be expressed, 
as in \eqref{definition Y pour NLIE} in terms of  $K_{\pm}$, $\mc{H}$. The latter determinants are parameterized by the zeroes $\paa{\tau_k}$
of $t_{\de}$, and the parameters $\paa{\de_k}$ appearing in the NLIE \eqref{TBA} coincide with the set of zeroes of the Hill determinant. 
By invoking the continuity of the logarithm on $\R$
and its decay at infinity, we get
\bem
k \Int{\R}{}  \f{\dd \mu}{2i\pi} \paa{ \pa{\mu+i\tf{\hbar}{2}}^{k-1}- \pa{\mu-i\tf{\hbar}{2}}^{k-1} }
\ln \pa{  1+ \f{ \rho^{\hbar} Y_{\de}\pa{\mu} }{ \abs{ \vth\pa{\mu-i\tf{\hbar}{2}} }^2 }   }  \\
= -\Int{\R-i\tf{\hbar}{2}}{}  \f{\dd \mu}{2i\pi} \paa{ \pa{\mu+i\hbar}^{k}- \mu^k }
\pac{ \f{K_+^{\prime}}{K_+}\pa{\mu} + \f{K_-^{\prime}}{K_-}\pa{\mu+i\hbar}  - \f{ \mc{H}^{\prime} }{ \mc{H} } \pa{\mu} } \\
=   \Int{ \substack{ \R+i\tf{\hbar}{2}  \rightarrow \\  \R-i\tf{\hbar}{2}  \leftarrow }  } {}  \f{\dd \mu}{2i\pi}  \mu^k
 \f{ \mc{H}^{\prime} }{ \mc{H} } \pa{\mu}   = \sul{p=1}{N} \pa{ \tau_p^k-\de_p^k } \;. \hspace{4cm}
\end{multline}

In the intermediate steps, we have used the quick decay at infinity of the integrand
\beq
\f{ K^{\prime}_{\pm} } { K_{\pm} }  \pa{\la}= \e{O}\pa{ \la^{-2N-1}}  \qquad \e{and} \qquad
 \f{ \mc{H}^{\prime} }{ \mc{H} } \pa{\la} = \e{O}\pa{\la^{-\infty}} \; .
\enq
This allows us to split the integral in three and compute the parts involving $K_{+}$, resp. $K_-$, by the residues in the upper/lower half plane (thus
giving 0).
Hence, the only part that gives a non-trivial contribution is the contour integral involving $\tf{\mc{H}^{\prime}}{\mc{H}}$.
The only poles that contribute to the result are located at the zeroes $\de_k$ of the Hill determinant (they have residue +1)
and at the poles $\tau_k$ of the Hill determinant (they have residue -1) that are located in the strip $\abs{\Im\pa{z}}< \tf{\hbar}{2}$. \qed

This result offers a direct way to recover the spectum of the 
model from a solution to the TBA equation \eqref{TBA}. 
It remains to derive the set of quantization conditions on the 
parameters $\de_k$.

\begin{thm}
\label{Theorem conditions de quantification}
There exists a unique entire solution $q$ to the T-Q equation \eqref{BAX} whose asymptotic
behavior is as stated in (ii) if and only if the parameters $\paa{\de_k}$ appearing in the TBA NLIE \eqref{TBA}
satisfy to the quantization conditions given in \eqref{ecriture conditions de quantification'}.
%
%
%
%
%
%
%
%
%
%
%
%

\end{thm}

\begin{rem}
The solvability of the quantization conditions, the occurrence of complex solutions ( $\Im\pa{\de_k} \not=0$,
$\de_k \in \paa{z \; : \; \abs{\Im\pa{z}}<\tf{\hbar}{2}}$, the uniqueness of solutions for a given choice of
integers $n_k \in \mathbb{Z}$ are all open questions.
\end{rem}

\proof

According to lemma \ref{lemme forme generale solutions TQ}, any meromorphic solution $q$ to the T-Q equation takes the form
\beq
q\pa{\la}=\f{ W\pac{q,Q_{\de}^{-}}\pa{\la}  }{ W\pac{Q_{\de}^+,Q_{\de}^-}\pa{\la}  } \cdot Q_{\de}^+\pa{\la}
- \f{ W\pac{q,Q_{\de}^+}\pa{\la}  }{ W\pac{Q_{\de}^+, Q_{\de}^-}\pa{\la}  } \cdot Q_{\de}^-\pa{\la} \; .
\enq
Recall that the Wronskian $W\pac{Q_{\de}^+,Q_{\de}^-}\pa{\la}$ is given by \eqref{q-Wronski}.
It is possible to compute the Wronskians $W\pac{q,Q_{\de}^{\pm}}\pa{\la}$ by using the asymptotic behavior of $q$ and $Q_{\de}^{\pm}$.
Due to their $i\hbar$ quasi-periodicity, these Wronskians take the form  $W\pac{q,Q_{\de}^{\pm}}\pa{\la}=\ex{-N\f{\pi}{\hbar}\la}\kappa^{-i\la} 
w_{\pm}\pa{\la}$, where $w_{\pm}\pa{\la}$ are entire
$i\hbar$-periodic functions. However, using the asymptotic behavior of $q$ and $Q_{\de}^{\pm}$
we get that $w_{\pm}\pa{\la}$ are bounded at infinity in the strip $\abs{\Im\pa{\la}}\leq\tf{\hbar}{2}$, and hence on $\Cx$. They are thus constant.
This proves the uniqueness of solutions for a given choice of $\tau_k$'s and hence $\de_k$'s. Indeed, up to a normalization constant, 
any solution $q$ satisfying to the requirements stated in point (ii), is of the form
\beq
q\pa{\la} =    \ex{\f{N\pi}{\hbar} \la} \f{  Q_{\de}^+\pa{\la}-\zeta Q_{\de}^-\pa{\la}  }{ \pl{k=1}{N}   \sinh\f{\pi}{\hbar}\pa{\la-\de_k}  }
\label{definition fonction q}
\enq
%
%
%
%
%
%
%
%
%
As the solution $q$ is entire, it has a vanishing residue at  $\la=\de_k$, $k=1\dots,N$. Therefore, the quantization conditions for the Toda chain appear as the set of $N-1$ conditions that
$q\pa{\la}$ has a vanishing residue at $\de_k$, $k=1,\dots,N$ supplemented with the $N^{\e{th}}$ quantization condition for the overall  momentum :
$\sum_{p=1}^{N}\tau_p=\sum_{p=1}^{N}\de_p=P$.
Note that it follows from $W\pac{Q_{\de}^+,Q_{\de}^-}\pa{\de_k+in\hbar}=0$ that if $q$ has a vanishing residue at a $\de_k$
then it also has a vanishing residue at $\de_k+i n \hbar$, $n\in \mathbb{Z}$. Therefore, there is indeed only a finite number $N$
of constraints of the parameters $\de$. The explicit form of these quantization conditions is then indeed as given in 
\eqref{ecriture conditions de quantification'}.

%
%
%
%
%
%
%
%
%
%

%
%
%
%
%
%
%
%
%
%
%
%
%

Conversely, if the quantization conditions are satisfied, then by taking $q$ as in \eqref{definition fonction q}, one 
obtains an entire solution with the desired asymptotics. \qed

\bibliographystyle{amsplain}
\bibliography{bibliotemple}

\end{document}